\documentclass{article}
\usepackage[a4paper, bindingoffset=0.5in, top=1in, bottom=1in, left=1in, right=1in, footskip=.25in]{geometry}
\usepackage[utf8]{inputenc}
\usepackage{graphicx}
\usepackage{multirow}
\usepackage{amsmath,amssymb,amsfonts}
\usepackage{mathrsfs}
\usepackage{hyperref}
\usepackage{url}
\usepackage{xcolor}

\newcommand{\SI}[2]{{{$#1 $}~{#2}}}

\newcommand{\celsius}{{$^\mathrm{\circ}$C}}

\date{}

\makeatletter
\renewcommand{\and}{%
  \end{tabular}\hskip 1em plus 0.5em minus 0.5em%
  \begin{tabular}[t]{c}}%
\makeatother

\begin{document}

\title{Nano-electronvolt Fourier-limited transition of a single surface-adsorbed molecule}

\author{
Masoud Mirzaei,\textsuperscript{1,2}\thanks{These authors contributed equally to this work.} \,
Alexey Shkarin,\textsuperscript{1}\textsuperscript{\thefootnote} \,
Burak Gurlek,\textsuperscript{3,1} \,
Johannes Zirkelbach,\textsuperscript{1,2,4} \\[0.3em]
Ashley J. Shin,\textsuperscript{1} \,
Irena Deperasi{\'n}ska,\textsuperscript{5} \,
Boleslaw Kozankiewicz,\textsuperscript{5} \,
Tobias Utikal,\textsuperscript{1} \\[0.3em]
Stephan G{\"o}tzinger,\textsuperscript{1,2,6} 
Vahid Sandoghdar\textsuperscript{1,2}\thanks{Corresponding author: vahid.sandoghdar@mpl.mpg.de}
}

\date{
\textsuperscript{1}Max Planck Institute for the Science of Light, 91058 Erlangen, Germany\\
\textsuperscript{2}Department of Physics, Friedrich Alexander University Erlangen-Nuremberg, 91058 Erlangen, Germany\\
\textsuperscript{3}Max Planck Institute for the Structure and Dynamics of Matter, 22761 Hamburg, Germany\\
\textsuperscript{4}Current Institution: Faculty of Physics, Ludwig-Maximilians-Universit{\"a}t M{\"u}nchen, 85748 Garching, Germany\\
\textsuperscript{5}Institute of Physics, Polish Academy of Sciences, 02-668 Warsaw, Poland\\
\textsuperscript{6}Graduate School in Advanced Optical Technologies (SAOT), Friedrich Alexander University Erlangen-Nuremberg, 91052 Erlangen, Germany
}

\maketitle

\section*{Abstract}
High-resolution spectroscopy allows one to probe weak interactions and to detect subtle phenomena. While such measurements are routinely performed on atoms and molecules in the gas phase, spectroscopy of adsorbed species on surfaces is faced with challenges. As a result, previous studies of surface-adsorbed molecules have fallen short of the ultimate resolution, where the transition linewidth is determined by the lifetime of the excited state. In this work, we conceive a new approach to surface deposition and report on Fourier-limited electronic transitions in single dibenzoterrylenes adsorbed onto the surface of an anthracene crystal. By performing spectroscopy and super-resolution microscopy at liquid helium temperature, we shed light on various properties of the adsorbed molecules. Our experimental results pave the way for a new class of experiments in surface science, where high spatial and spectral resolution can be combined.

\section*{Introduction}
Textbook physics usually treats surfaces as mathematical entities that are smooth in two dimensions and manifest a sharp boundary between two media. In reality, however, surfaces are governed by atomic and molecular textures, which become specially consequential in their immediate vicinity, where adsorbed species reside. Beautiful experiments have visualized the contact between individual physisorbed molecules and the outermost atoms of surfaces by using scanning tunnelling microscopy (STM), atomic force microscopy (AFM), scanning near-field optical microscopy (SNOM) and scanning transmission electron microscopy (STEM) \cite{Chiang-Science-14, Iwata-NatComm-15, Yang-NatPhoton-20, Kharel-NL-22}. A quantum mechanical understanding of the adsorbed entities, however, would also require high-resolution information about their energy levels. These investigations have remained elusive, as the spectra of adsorbed atoms and molecules have been broadened by environmental effects.

\begin{figure}
\centering
\includegraphics[width=90mm]{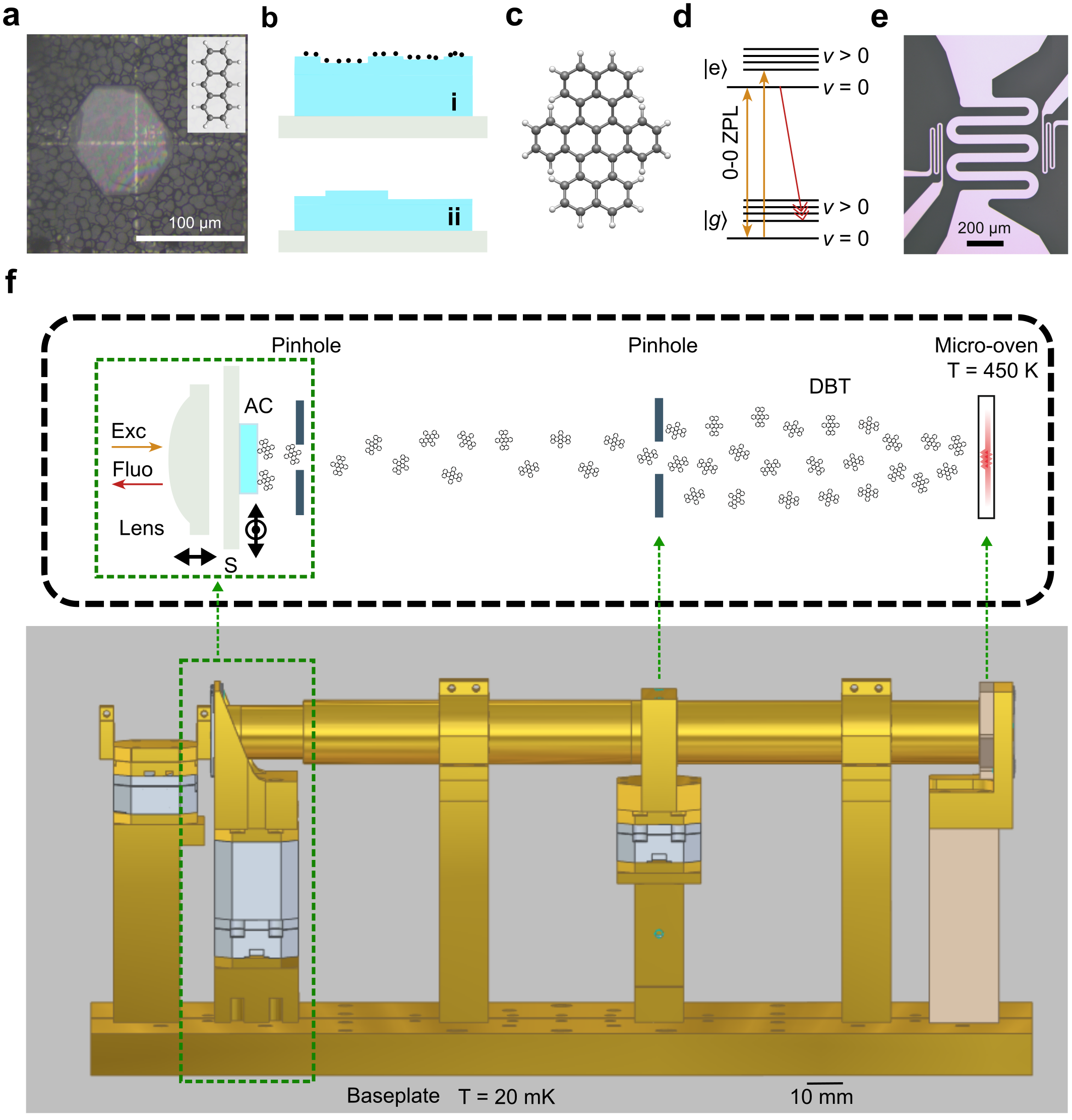}
\caption{\textbf{Sample preparation and experimental setup.} \textbf{a,} Microscope image of a large AC crystal surrounded by many small crystallites, which were sublimated onto a cover glass. Inset shows the structure of AC. Carbon and hydrogen atoms are shown in dark and light shades, respectively. \textbf{b,} Sketch of the crystal self-cleaning process. (i) After original sublimation and exposure to air, the crystal may contain surface contaminants, indicated by the black dots. (ii) Partial sublimation in vacuum results in a thinner crystal with a fresh surface. \textbf{c,} Molecular structure of DBT. \textbf{d,} Jablonski energy diagram of DBT. $\vert g \rangle$ and $\vert e \rangle$ stand for the electronic ground and excited states, respectively, while v denotes the order of the vibrational levels in each state. \textbf{e,} Microscope image of the custom-designed oven chip containing a heater (central part) and Pt thermometers (sides). \textbf{f,} Experimental setup for depositing DBT and achieving optical access for spectroscopy and microscopy. Upper panel: schematic layout; lower panel: side view to scale. The micro-oven evaporates DBT onto a fresh AC crystal surface located on a glass substrate (S). Two pinholes define the molecular beam. The oven chip is thermally isolated by a support made of polyetheretherketone (light brown). The other components are made of gold-coated oxygen-free copper. The path connecting the oven chip and the crystal surface is enclosed within a tube to avoid contamination of the cryostat. An aspheric lens is used for laser illumination and collection of fluorescence from the sample. Piezo-driven stages (gray) allow for the translation of the sample, the aspheric lens, and one of the pinholes. }
\label{fig:setup}
\end{figure}

Rigorous surface studies confront the challenge that surfaces easily accumulate defects and are polluted by adsorption of unwanted molecules. Everyday surfaces, for instance, can be covered by several monolayers of water as well as large populations of carbon-rich molecules under ambient conditions \cite{asay2005}. Surface scientists usually address this issue by using pristine surfaces prepared \textit{in situ} under ultrahigh vacuum ($P \leq10^{-10}$\,bar) \cite{Surface-Physics-book-Fauster-Schneider}. Following this procedure, STM architectures and plasmonic effects were exploited to perform electro- and photoluminescence spectroscopy of single molecules physisorbed on surfaces at cryogenic temperatures \cite{Chong-NatComm-16, Zhang-Nature-13, Imada-Science-20}. The narrowest reported linewidth in these experiments was obtained from molecules on a thin sodium chloride crystal and has been limited to about 500\,$\mu$eV, possibly due to the presence of metallic tips \cite{Imada-Science-20}. Direct laser spectroscopy on a similar substrate yielded considerably lower linewidths in the order of 1\,$\mu$eV, equivalent to 450\,MHz \cite{marquardt2021}. Other direct laser spectroscopy efforts, including a recent single-molecule work on a hexagonal boron-nitride have also reported $\mu$eV linewidths \cite{Basche-ChemPhysLett-91, Smit-NatComm-23}, one order of magnitude larger than the expected Fourier limit. 

\section*{Experimental approach}
In this work, we introduce a new approach for producing fresh surfaces of anthracene (AC) crystals at cryogenic conditions and study single dibenzoterrylene (DBT) molecules deposited on them \textit{in situ}. The central features of our strategy for sample preparation are depicted in Fig.\,\ref{fig:setup}. We start with AC crystals that were sublimated onto a coverglass. An example is shown in Fig.\,\ref{fig:setup}a, where many small crystallites surround a larger hexagonal crystal with a thickness of a few tens of micrometers. The results in this work were obtained on such large crystals.

The crystal sample is transferred to a closed-cycle dilution cryostat, where the sample temperature changes from about \SI{20}{\celsius} to \SI{-5}{\celsius} in three hours under vacuum ($P< $\SI{10^{-4}}{mbar}). During this time, the outer layer of the AC crystal sublimates at a rate of few micrometers per hour although the process is substantially slowed down as the temperature drops. The sublimation process leaves behind a surface with molecularly flat terraces connected by monolayer steps (Fig.\,\ref{fig:setup}b), as verified by AFM measurements. After cooling to liquid helium temperature, the crystal can be preserved indefinitely and its surface is fully protected against contamination because the vapor pressure of all materials becomes negligible \cite{pobell1996}. The combination of high vacuum and cryogenic temperatures also strongly suppresses re-adsorption of residual atoms and molecules in the cool-down period. However, not having the means for \textit{in situ} characterization of the surface, we cannot assess its cleanliness in a quantitative manner. 

DBT molecules (Fig.\,\ref{fig:setup}c,d) were evaporated onto the freshly prepared AC crystal surface \textit{in situ} at \SI{4}{K}. To minimize the thermal load in the cryostat, we fabricated a chip-based oven consisting of a meandering platinum micro-wire on a glass substrate, covered with a protective layer of silica (Fig.\,\ref{fig:setup}e, see also Methods). DBT powder was distributed over the heater wire area before the oven chip was inserted into the cryostat. Figure\,\ref{fig:setup}{f} shows the schematics of the experimental rail system in the dilution cryostat. By applying  \SI{10}{V} to the heating micro-wire, corresponding to a dissipated power of about \SI{0.5}{W}, we locally raised the oven temperature to approximately \SI{450}{K} for evaporating DBT molecules, while the sample temperature remained below 6\,K throughout the evaporation process. After one minute, we turned off the oven and let the whole setup cool down to \SI{20}{mK}. 

We monitored the surface coverage of DBT via its fluorescence signal during the deposition procedure by exciting an ensemble of molecules to a higher-lying vibrational state of the electronic excited state (see Fig.\,\ref{fig:setup}d). Laser illumination and fluorescence collection were carried out through the AC crystal using an aspheric lens. The details of the optical system can be found in the previous publication \cite{Zirkelbach2023}. 

\begin{figure*}
\centering
\includegraphics[width=130mm]{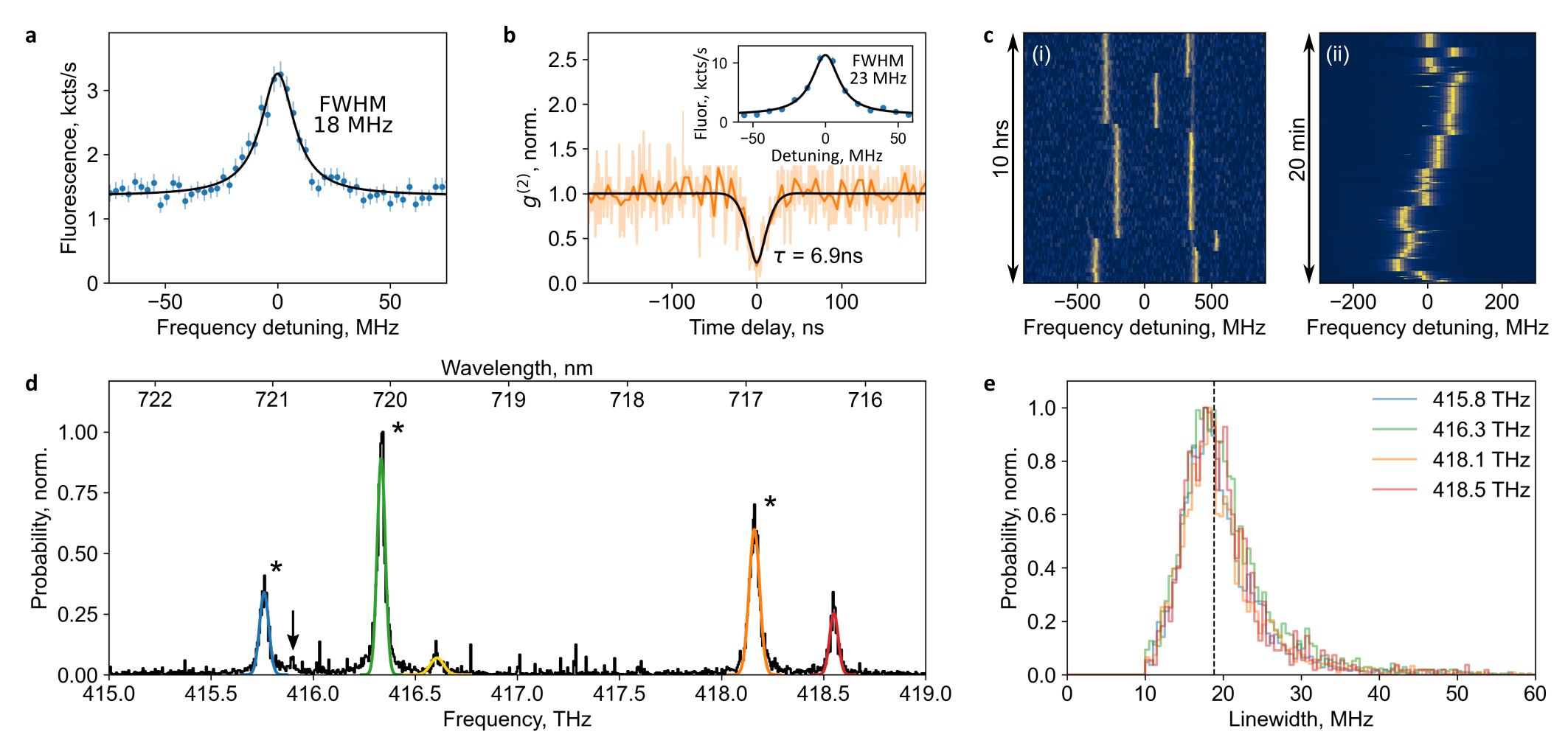}
\caption{\textbf{Spectral and temporal properties of individual DBT molecules.} \textbf{a,}
Fluorescence excitation spectrum of a DBT molecule on an AC surface. Solid line shows a Lorentzian fit function with FWHM of 18\,MHz. \textbf{b,} Second-order autocorrelation function of the emitted photons from another single molecule. Inset: a low-power fluorescence excitation spectrum of the same molecule. \textbf{c,} (i) Repeated laser frequency scans at low excitation power of \SI{8}{nW} performed once every 10 minutes. 00ZPLs remain stable over several hours. (ii) Repeated laser frequency scans performed once every two seconds at higher excitation power of \SI{150}{nW} result in spectral instability. \textbf{d,} Distribution of 00ZPLs from about 50,000 molecules show distinct spectral sites with central frequencies and FWHM linewidths ($\nu_i, \Delta\nu_i$): $\nu_1=415.8$\,THz, $\Delta\nu_1=68$\,GHz; $\nu_2=416.3$\,THz, $\Delta\nu_2=56$\,GHz; $\nu_3=416.6$\,THz, $\Delta\nu_3=65$\,GHz; $\nu_4=418.1$\,THz, $\Delta\nu_4=63$\,GHz; $\nu_5=418.5$\,THz, $\Delta\nu_5=59$\,GHz. The arrow marks a possible sixth site. The sites with long-term stability are marked by an asterisk. \textbf{e,} Distribution of the 00ZPL linewidths from about 10,000 single molecules for the four most prominent sites with the same color code as in (d). The dashed line denotes the median linewidth at 19\,MHz.}
\label{fig:Fig2_IHB_jumps_g2}
\end{figure*}

Once the sample was prepared, we investigated the DBT molecules via fluorescence excitation spectroscopy of their zero-phonon lines connecting the lowest vibrational levels (v=0) of the electronic ground and first excited states, referred to as 00ZPL (Fig.\,\ref{fig:setup}d). By scanning the frequency of a continuous-wave laser (linewidth $<$ 1\,MHz), we could selectively excite single molecules because the narrowing of 00ZPL transitions at low temperatures prevents the overlap of the resonances from a large number of molecules in the observation volume  \cite{SM-book1996}. 

\section*{Results}
Figure\,\ref{fig:Fig2_IHB_jumps_g2}{a} displays an exemplary fluorescence excitation spectrum of the 00ZPL from a single DBT molecule on an AC crystal surface recorded at the wavelength of 720\,nm with an excitation power well below saturation. The full width at half-maximum (FWHM) of the Lorentzian resonance corresponds to $\Gamma=18$\,MHz, which is significantly lower than the values of 27-\SI{40}{MHz} reported for bulk DBT:AC \cite{nicolet2007, hofmann2005}. While narrow isolated spectra in a dilute sample are usually strong indicators that the signal stems from a single emitter, an unequivocal proof requires antibunching measurements \cite{SM-book1996}. Figure\,\ref{fig:Fig2_IHB_jumps_g2}{b} shows an example of the second-order autocorrelation function for the fluorescence signal from another molecule, where an antibunching dip well below 0.5 at zero time delay testifies to the signal stemming from an individual quantum emitter. Moreover, the rise time reports on the excited-state lifetime, $\tau=6.9$\,ns, of this molecule. This value matches the measured 00ZPL linewidth of $\Gamma=23$\,MHz (inset, Fig.\,\ref{fig:Fig2_IHB_jumps_g2}{b}) according to the relationship $\Gamma=1/(2\pi\tau)$ for a Fourier-limited transition. We will address the slight variations in the observed linewidths later.
 
Having demonstrated narrow homogeneous linewidths, we now examine the phenomena of spectral diffusion and jumps, as are commonly observed in solid-state emitters \cite{Orrit-review-2000,basche-spectral-diffusion-1992, Orrit2009, aharonovich2016a}. Figure\,\ref{fig:Fig2_IHB_jumps_g2}{c} plots examples of long-term spectral stability, demonstrating center frequency deviations that can be as small as 0.1\,$\Gamma$ over several hours under weak excitation (i). Panel (ii) of this figure, however, shows that frequent spectral jumps can occur under high excitation intensities. Such light-enhanced instabilities are likely caused by energy transfer from the excited DBT molecule to its nearby AC molecules, thus, modifying the environment \cite{Orrit-review-2000}. 

To obtain a comprehensive view of the inhomogeneous broadening (IHB), we scanned a very wide spectral region between \SI{700}{nm} to \SI{789}{nm}. Figure\,\ref{fig:Fig2_IHB_jumps_g2}d displays that the 00ZPL frequency of DBT can take on different values. We identify five well-defined spectral sites with a possibly sixth site marked by an arrow. While the sites at 415.8, 416.3 and 418.1\,THz were stable for months, the other sites disappeared after several days following deposition, which we attribute to photo-induced processes. As a result, the data presented in this work mostly stem from the three persistent sites, marked by asterisks in Fig.\,\ref{fig:Fig2_IHB_jumps_g2}d. A noteworthy feature of all IHB sites is their low FWHM, in the order of 60\,GHz, which is about twice smaller than the width of the IHB for bulk DBT:AC \cite{nicolet2007} and about two orders of magnitude smaller than a recently reported spectrum of a PAH molecule on a surface \cite{Smit-NatComm-23}. 

In Fig.\,\ref{fig:Fig2_IHB_jumps_g2}{e}, we also present the 00ZPL linewidths recorded from single molecules of the four dominant spectral sites. Remarkably, all sites reveal a narrow distribution with a median at 19\,MHz. The low widths of the IHB sites and of the single-molecule 00ZPLs present a very strong evidence for a high degree of order in the way DBT molecules are positioned against the crystal lattice sites of the AC surface. We note that these properties are in strong contrast to the GHz homogeneous linewidths and IHB of up to \SI{10}{THz} known from guest molecules embedded in disordered systems like polymers \cite{orrit1992}.

\begin{figure}
\centering
\includegraphics[width=90mm]{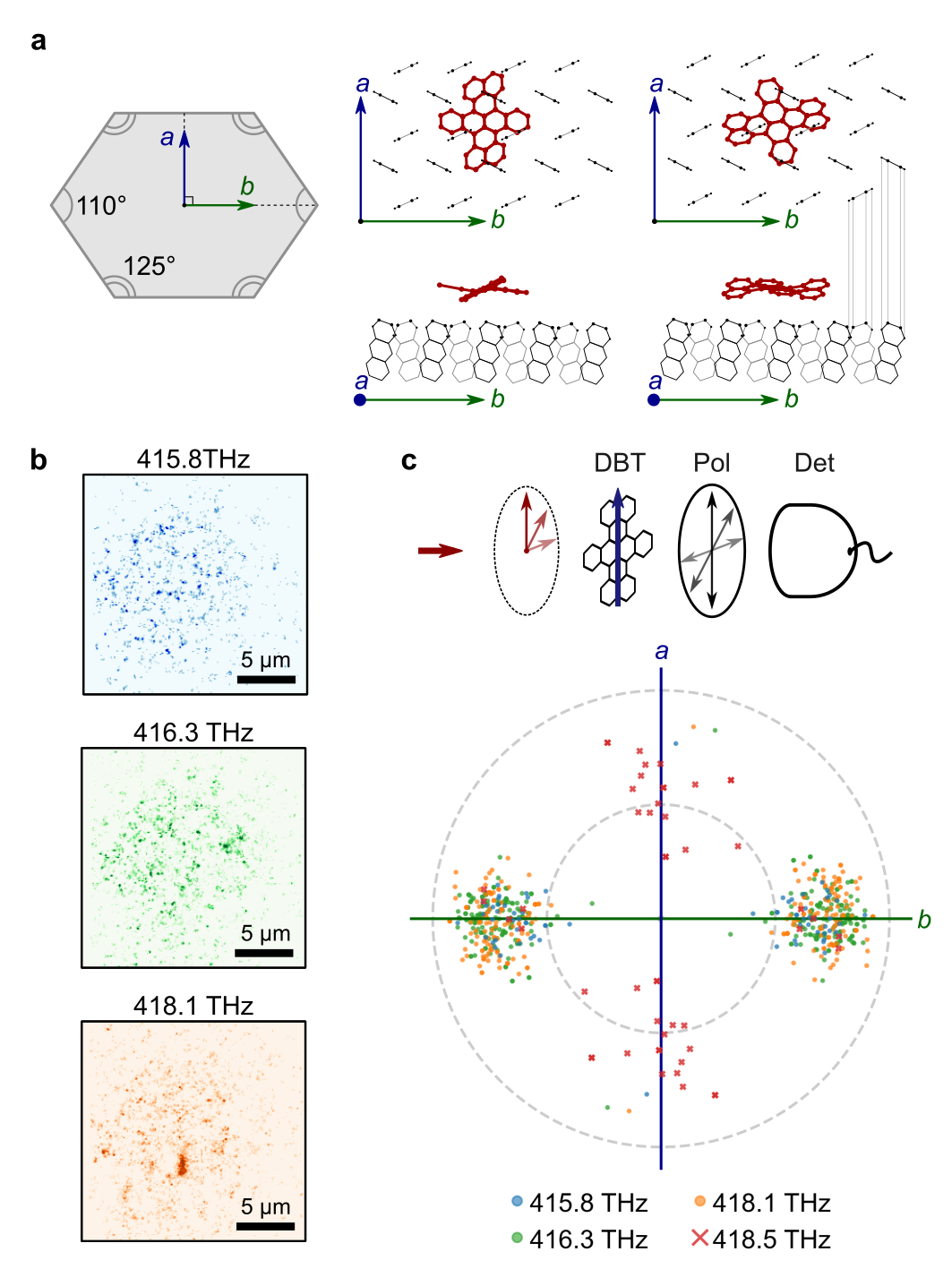}
\caption{\textbf{Spatial and orientational distribution of DBT molecules on the AC surface.} \textbf{a,} (Left) Top-view geometry of an AC crystal and the orientations of its \textit{a} and \textit{b} axes. (Right) Theoretically predicted configuration of the carbon atoms (red) of a DBT molecule with respect to the carbon atoms (black) of the AC crystal. Two orientations are found for the adsorption of DBT. (Upper row) View from above, i.e., in the \textit{ab} plane. (Lower row) View from the side. \textbf{b,} Super-resolution maps of individual DBT molecules for the three persistent spectroscopic sites. \textbf{c,} (Top) Schematics for polarization analysis of the fluorescence signal after excitation with linearly-polarized light at different angles within $180^\circ$. (Bottom) Pairs of symbols mirrored through the origin of the polar plot represent the orientation of a single molecule from one of the spectral sites as marked in the legend. The distance of a point from the center corresponds to the degree of polarization, i.e., visibility $V$ of the fluorescence modulation. Dashed circles mark $V=0.5$ and $V=1$.}\label{fig:localization_polarization}
\end{figure}

To shed light on the origin of the different spectral sites and the underlying order in DBT adsorption, we carried out several theoretical and experimental investigations. Figure\,\ref{fig:localization_polarization}{a} shows a schematic of the AC crystal geometry and the results of hybrid quantum mechanical and molecular dynamics calculations using the ONIOM method \cite{Chung-ChemRev-15} optimized for the adsorption of DBT on the AC crystal surface (Methods). The outcome shows that DBT displays two orientations roughly along the \textit{a} and \textit{b} axes of the AC crystal. In addition, previous theoretical studies predicted that isolated DBT molecules can possess three stable conformational isomers \cite{Deperasiska2010}. The combination of these findings offers a plausible scenario for six different interaction energies between DBT and the AC crystal. 

Experimental insight into the exact configuration of DBT against the AC lattice at the surface can help elucidate the nature of the spectral sites. Previous studies of adsorbates have used scanning probe microscopy to map adsorbate atoms and molecules with respect to the outer atoms of a substrate surface with exquisite spatial resolution \cite{Iwata-NatComm-15,Pengcheng-NatComm-2023,brown2024}. Although we did not have access to such a high degree of spatial information, narrow 00ZPL resonances allowed us to identify and localize single DBT molecules with a precision of 50\,nm. Figure\,\ref{fig:localization_polarization}{b} presents super-resolution spatial maps of DBT molecules for the three persistent spectroscopic sites. In all cases, we find a fairly uniform distribution of molecules over tens of microns. We, thus, conclude that local terraces and steps cannot be responsible for the emergence of various spectral sites. 

Next, we determined the orientation of the individual DBT molecules by analyzing the polarization of their emission upon excitation with linearly polarized light at various angles \cite{nicolet2007b}. Figure\,\ref{fig:localization_polarization}{c} shows the outcome for molecules in the four dominant spectral sites presented in Fig.\,\ref{fig:Fig2_IHB_jumps_g2}{d}. Here, the fluorescence signal from each molecule was analyzed to determine the visibility of the fluorescence modulation and to extract the angle corresponding to the maximal signal. Within the accuracy of our measurements, we found the molecules in the three persistent sites to align with the \textit{b}-axis of the AC crystal whereas molecules in the fourth site were aligned along the crystal \textit{a}-axis. This supports the above-mentioned theoretical prediction that DBT orients itself along anthracene's two crystal axes. 

\begin{figure}
\centering
\includegraphics[width=90mm]{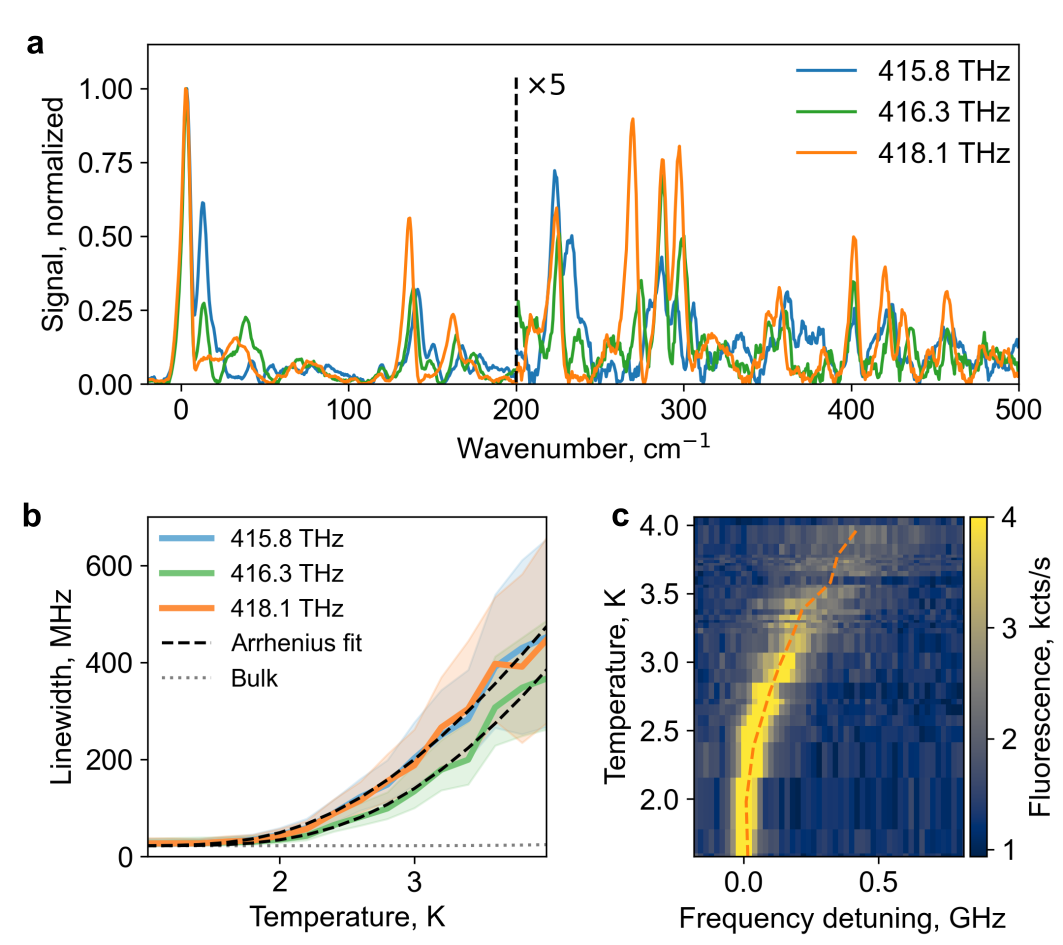}
\caption{\textbf{Emission spectra and temperature dependence of 00ZPL.} \textbf{a,} 
Fluorescence spectra recorded from ensembles of molecules excited to higher-lying vibrational levels of the electronic excited state in the range of $420-423$\,THz. The 00ZPL is set at the origin of the frequency axis in each case. Colors indicate data from different persistent spectral sites. The signal is represented in x5 magnification for the region beyond the dashed line. \textbf{b,} 00ZPL linewidth as a function of temperature obtained from about 100 molecules for the three different spectral sites. Dashed curves show Arrhenius fits. For comparison, the dotted curve shows the behavior of DBT in bulk AC \cite{nicolet2007b}. \textbf{c,} Evolution of the 00ZPL spectrum of a single molecule as a function of temperature. The orange dashed curve marks the 00ZPL center in each laser frequency scan.}
\label{fig:Spectra_Tdeph}
\end{figure}

To explore the properties of different spectral sites further, we recorded emission spectra from DBT molecules of the three persistent spectral sites. Here, we excited the molecules via higher-lying vibrational states of the excited state (see Fig.\,\ref{fig:setup}{d}) and analyzed their red-shifted fluorescence using a grating spectrometer. Figure\,\ref{fig:Spectra_Tdeph}a presents the results obtained from ensembles of molecules for robust statistics. One recognizes systematic differences among the spectra of the molecules from the different sites. In particular, we observe Lorentzian-like peaks in the region that immediately follows the 00ZPLs. These phonon wings stem from the coupling of a guest molecule to the phonon modes of the matrix. The sharp features, which are especially pronounced for the sites at 415.8 and 416.3\,THz, do not appear in the bulk DBT:AC emission spectra \cite{clear2020}. They are representative of pseudo-local phonon modes \cite{Hsu-psu-1985,gurlek-2025} and encode information about the way DBT interacts with the AC surface. Since all molecules in the three spectral sites share a common orientation along the crystal \textit{b} axis, their different emission spectra may hint at the role of the three possible conformational isomers of DBT \cite{Deperasiska2010}. 

Above, we examined the vibronic transitions associated with the intramolecular motion of DBT as well as the 00ZPL phonon wings, which report on the coupling of a DBT molecule with its AC environment. We now turn our attention to the temperature dependence of the 00ZPL. In general, electronic transitions of quantum emitters embedded in a solid matrix can be broadened by dephasing and be shifted in energy by electron-phonon interactions~\cite{Skinner1988}. In Fig.\,\ref{fig:Spectra_Tdeph}b, we plot the change of the linewidth as a function of temperature assessed from about 100 DBT molecules for the three persistent spectral sites. Notably, dephasing of the 00ZPL already becomes appreciable at around \SI{2}{K}, incurring linewidth broadenings that reach several hundreds of MHz at \SI{4}{K}. The dashed curves in Fig.\,\ref{fig:Spectra_Tdeph}b show that the behavior of the linewidth ($\Gamma_2^{*}$) agrees with an Arrhenius trend given by $ \Gamma_2^{*}(T) = A \exp(-T_A/T) $, where $T_A$ denotes an activation temperature~\cite{nicolet2007}. For comparison, the dotted line shows that DBT in bulk AC experiences negligible dephasing up to \SI{4}{K} \cite{nicolet2007b}. 

It is known that the local phonon density of states is increased at a surface compared to that of bulk \cite{Geller2004, Tighineanu-PRL-2018}. We believe this effect can explain the observed lower activation temperature for physisorbed molecules if one considers pure dephasing theories based on acoustic phonons~\cite{Skinner1988,clear2020}. Figure\,\ref{fig:Spectra_Tdeph}{c} depicts an exemplary modification of the 00ZPL for a single molecule as the temperature is raised to 4\,K. The observed frequency shift matches the predictions of an analytical model that incorporates pseudo-local phonons and uses the experimentally measured linewidth as an input~\cite{Hsu-exp-1985}. No parameters in the model were adjusted to fit the experimental data. Our results motivate future dedicated studies of phonon-mediated spectral changes at various surfaces, including thin films and engineered devices.

\begin{figure}
\centering
\includegraphics[width=90mm]{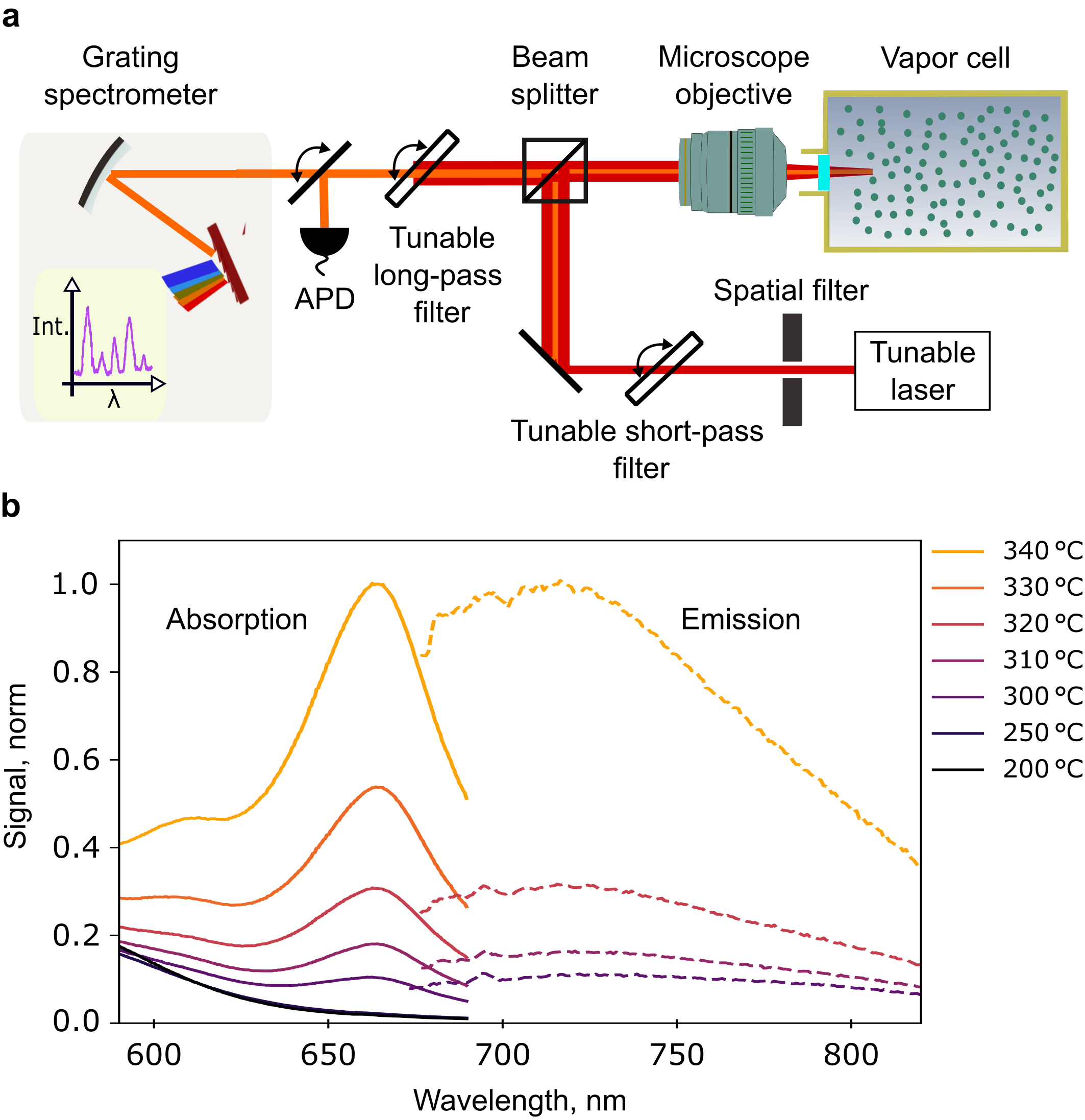}
\caption{\textbf{Spectroscopy of DBT in a vapor cell.} \textbf{a,} Schematics of the vapor cell measurement setup. DBT is heated in a cell that is prepared in a glove box under nitrogen at atmospheric pressure. A sapphire window on the cell and a long working distance microscope objective provide optical access for laser spectroscopy (Methods). \textbf{b,} Absorption (solid) and emission (dashed) spectra of DBT recorded at different oven temperatures noted in the legend. Not all temperatures have associated emission data. The spectra in the two categories are normalized by the maxima of their corresponding data recorded at \SI{340}{\celsius}. Control measurements confirmed that the rise of the signal at 600\,nm is also present in a cell without DBT (i.e., empty), while the absorption peak at around 680\,nm arises exclusively from DBT\iffalse (Supplementary Note S12)\fi. The high number of vibrational and rotational degrees of freedom in the gas phase results in broad spectra. The estimated Doppler broadening of 0.35\,GHz and collisional broadening below 5\,GHz are both negligible on the scale of the presented spectra\iffalse (Supplementary Note S13)\fi. }
\label{fig:vapor}
\end{figure}

As a last topic, we consider the observed transition frequency for surface adsorbed DBT molecules. Physisorbed molecules reside at the minimum of a Lennard-Jones potential which is dominated by a repulsive component in the contact region and an attractive van der Waals contribution at larger distances away from the surface \cite{LennardJones1932}. While the van der Waals potential has been successfully studied for both ground and excited states of atoms at separations of tens to hundreds of nanometers \cite{Sandoghdar1992,Peyrot2019}, high-resolution studies of the full Lennard-Jones energies for adsorbed species have remained elusive. On the theoretical side, DFT calculations for simple systems have made great progress over the past two decades \cite{Chakarova2006,Maier2022}, but they are still challenging for arbitrary surfaces. Experimental difficulties include the realization of well-defined surfaces, control and precise characterization of the distance between an emitter and a surface, as well as suitable methods for precision spectroscopy. Fourier-limited transitions of single physisorbed molecules would provide a unique opportunity for high-resolution investigation of frequency shifts caused by the presence of the AC crystal surface. To pursue this goal, we set out to determine the 00ZPL of unperturbed DBT by performing spectroscopy in the vapor phase. This information had not been reported in the literature. 

The low vapor pressures of large PAHs pose a formidable challenge for the realization of a gas cell because it requires high temperatures\,\cite{ruiterkamp2002}, so we devised a dedicated setup as depicted in Fig.\,\ref{fig:vapor}a (Methods). Figure\,\ref{fig:vapor}{b} displays spectra recorded from DBT in a nitrogen buffer gas at different oven temperatures. To record absorption spectra, we scanned the frequency of the excitation laser beam and collected the total fluorescence signal on an avalanche photodiode. Emission spectra were obtained by exciting the molecules at the fixed laser wavelength of 650\,nm and detecting their fluorescence using a grating spectrometer.

The crossing of the absorption and emission spectra at wavelength $\lambda=682$\,nm, equivalent to the frequency of 440\,THz, can be attributed to the position of the 00ZPL for isolated DBT. We, thus, assess a frequency shift in the range of 22-25\,THz for the different spectral sites of DBT adsorbed on the AC crystal surface. Interestingly, an analytical model for the attractive van der Waals energy between a point dipole and a dielectric surface \cite{Courtois1996} yields a frequency shift of 24\,THz for the parameters of our system. The agreement with our measurement is intriguing since this theoretical model neglects subtleties such as the validity of the dipole approximation and is based on the description of the interface by the macroscopic concept of refractive index. A robust theoretical treatment that takes into account the molecular structures of DBT and the AC surface is beyond the scope of our work. However, we believe  precision spectroscopy introduced in this work will motivate quantitative comparisons between high-resolution theoretical and experimental studies to probe fine details of molecular interactions.

\section*{Discussion and Outlook}Fourier-limited electronic transitions of molecules were previously only detected in bulk organic crystals \cite{toninelli2021a, adhikari2022}. Using the adsorption of dibenzoterrylene on an anthracene crystal as a case study, we have now shown that this degree of spectral resolution is also within reach on surfaces. The achieved resolution surpasses the reports from STM experiments by more than a factor of 5000 \cite{Imada-Science-20} and is better than previous laser spectroscopy results of other adsorbate-surface systems by a factor of 30 \cite{marquardt2021, Smit-NatComm-23}. We postulate that contrary to the previous findings \cite{Basche-ChemPhysLett-91, Fleury-JCP, Wang-APL-04, Liu-PRAppl-18, Vainer-Faraday-15, marquardt2021, Smit-NatComm-23} and the common belief that the vicinity of surfaces causes broad lines, a suitable choice of material and proper preparation of its surface can lead to Fourier-limited spectra. 

Extension of our work will help shed light on the origin of the spectral sites of DBT and its emission spectra in conjunction with the way DBT molecules sit against the AC surface atoms. Our experimental findings are consistent with the possibility that the three stable spectral sites would be attributed to the alignment of the three isomer configurations of DBT along the \textit{b}-axis of the AC crystal. Building on this logic, we believe that the three unstable sites might also be associated with DBT's three isomeric conformations aligned with the \textit{a}-axis. Indeed, the observation that DBT in one of these sites is oriented in the direction of the \textit{a}-axis strengthens this hypothesis. Furthermore, support for the stability of DBT aligned along the \textit{b}-axis comes from ONIOM calculations, which reveal a systematically lower adsorption energy than for molecules oriented in the \textit{a} direction. More experimental and theoretical studies are needed to explore and verify these hypotheses.

We anticipate that future research will identify other adsorbates and surfaces with which the Fourier limit can be reached. For instance, investigation of molecules such as porphyrins, phthalocyanins and various organometallic compounds will be of great interest, as they not only play an enormous role in biology \cite{Nikolaou-Frontiers-2024} but are also being explored for quantum technology \cite{Bayliss-Science-2020, shin2024}. Indeed, reports of MHz (neV) linewidths of 00ZPLs from single Mg Tetra-azaporphyrins in bulk \cite{Starukhin-2001} suggest that Fourier-limited resonances for this system might also be within reach on surfaces. 

A particularly attractive future research direction would be to build on the progress in STM \cite{Chiang-Science-14, Chong-NatComm-16,Imada-Science-20} and AFM \cite{Gross-Science-2009,Pengcheng-NatComm-2023} studies to combine the spatial resolution of scanning probe microscopy with Fourier-limited resolution in laser spectroscopy. While in STM the metallic tip and the underlying substrate can lead to quenching and line broadening, the use of dielectric tips functionalized with CO molecules in non-contact mode AFM \cite{swart-ChemCommun-2013, Fang-Matter-2021} offers a promising platform for such investigations. It would be intriguing to use cryogenic scanning probe microscopy \cite{Michaelis-Nature-2000, Gerhardt-PRL-2007} to manipulate and image individual bonds and atoms \cite{Pengcheng-NatComm-2021} while monitoring the electronic and vibronic transitions of single molecules \cite{Chen-PRL-2010,Kong-NatComm-2021} with nano-electronvolt spectral resolution. Precision spectroscopy of adsorbates promises to provide direct insight into their quantum mechanical properties with immediate implications for surface science, chemistry and quantum technology. 




\section*{Methods}

\subsubsection*{Cryogenic micro-oven} The micro-oven shown in Fig.\,\ref{fig:setup}e consists of a meandering heating wire with a width of \SI{100}{\textmu m} and thickness of \SI{100}{nm}. It covers an area of \SI{1}{$\mathrm{mm^2}$}, resulting in a room-temperature resistance of about \SI{90}{Ohm}. Platinum was chosen due to the strong and well-characterized temperature dependence of its resistance, which allowed us to estimate the temperature of the heater during the operation. The DBT molecules are deposited on top of the slide directly in their powder form and stay attached thanks to the electrostatic attraction. The micro-oven is placed on a holder at about \SI{20}{cm} from the sample.  

\subsubsection*{Vapor cell} To prepare DBT molecules in the vapor phase, we used a tube furnace (type LOSA from HTM Reetz GmbH) capable of reaching temperatures of \SI{350}{\celsius}. The furnace housed a custom-made stainless steel cell with a sapphire window for optical access. About \SI{0.4}{mg} of DBT powder was placed in the cell inside a glove box under nitrogen atmosphere ($<$\SI{30}{ppm} $\mathrm{H_2O}$, $<$\SI{300}{ppm} $\mathrm{O_2}$), and the cell was subsequently sealed to maintain a chemically inert environment.

\subsubsection*{Optical setup for gas phase measurements}The optical setup used to investigate the spectroscopic properties of DBT molecules in the gas phase is shown in Fig.\,\ref{fig:vapor}a. We used a pulsed supercontinuum laser equipped with a narrow-band (\SI{2}{nm}) filter tunable within the wavelength range between 500 and \SI{700}{nm}. After passing through two tunable short-pass filters (cut-off wavelength of between \SI{620}{nm} and \SI{710}{nm}) to clean the spectrum, the laser was focused inside the cell \SI{2}{mm} past the entrance window using an objective with a numerical aperture of 0.26 and a working distance of \SI{31}{mm}. The fluorescence emitted from the cell was captured by the same objective and directed towards the detection path, where two fixed long-pass filters (cut-off wavelength of \SI{705}{nm}) eliminated any remaining excitation light before the fluorescence was detected by an avalanche photodiode (measurements of the excitation spectrum) or directed towards a grating spectrometer using a multi-mode fiber (measurements of the emission spectrum). The excitation spectrum was normalized by the laser power in front of the objective, which was measured separately.

\subsubsection*{Surface topography}
To estimate the sublimation rate of AC crystals under vacuum, samples were placed in a test chamber evacuated to \SI{10^{-6}}{mbar} at room temperature. Changes in crystal thickness during sublimation were monitored via optical interference, as color variations corresponded to shifts in thickness. The rate was determined by tracking the evolution of interference colors over time, accounting for the refractive indices. The sublimation rate was estimated to be approximately $2\,\mu$m/hr. After one hour of vacuum exposure, the crystal surfaces were characterized using tapping-mode AFM, which revealed step-like topographic features with heights of approximately 1\,nm and lateral dimensions ranging from sub-micron to several microns.

\subsubsection*{Ab initio calculations}
The structure of DBT adsorbed on an anthracene surface was optimized using the ONIOM method as implemented in Gaussian~\cite{frisch2016gaussian}. The DBT molecule was treated at the DFT level with the B3LYP functional~\cite{Becke1988,Lee1988} and 6-31G(d,p) basis set, while the anthracene monolayer (45 molecules arranged as in the bulk crystal) was described using the Universal Force Field. Multiple initial orientations of DBT were considered.

The vertical excitation energy of DBT in the gas phase was computed at the STEOM-DLPNO-CCSD level~\cite{BerraudPache2019} with TightPNO settings and a def2-TZVP(-f) basis, using a geometry optimized at the wB97X-D3/def2-TZVP(-f) level, yielding 698~nm.




\subsubsection*{Supplementary information}

Supplementary materials are not included.

\subsubsection*{Acknowledgements}

We thank Maksim Schwab for valuable advice during the design phase and building the cryogenic components of the setup. We are also grateful to Oliver Bittel and Lothar Meier for their assistance with the electronics. We thank Jan Renger, Isabel G\"assner, and Irina Harder for the fabrication of the micro-oven chip. I.D. and B.K. acknowledge the Interdisciplinary Centre for Mathematical and Computational Modelling (ICM) of the University of Warsaw for performing calculations under the computational allocation G98-2100. This work was funded by the Max Planck Society.







\end{document}